\documentclass{elsart}

\usepackage{epsfig}

\usepackage{amssymb}

\newcommand{\bnl}           {$\rm^{1}$}
\newcommand{\ires}          {$\rm^{2}$}
\newcommand{\kraknuc}       {$\rm^{3}$}
\newcommand{\krakow}        {$\rm^{4}$}
\newcommand{\baltimore}     {$\rm^{5}$}
\newcommand{\newyork}       {$\rm^{6}$}
\newcommand{\nbi}           {$\rm^{7}$}
\newcommand{\texas}         {$\rm^{8}$}
\newcommand{\bergen}        {$\rm^{9}$}
\newcommand{\bergenc}       {$\rm^{9*}$}
\newcommand{\bucharest}     {$\rm^{10}$}
\newcommand{\kansas}        {$\rm^{11}$}
\newcommand{\oslo}          {$\rm^{12}$}

\begin{document}

\begin{frontmatter}

\title{Nuclear Modification Factor for Charged Pions and Protons at Forward Rapidity in Central Au+Au Collisions at 200 GeV}


\begin{center}
\author{
  I.~Arsene\bucharest,
  I.~G.~Bearden\nbi, 
  D.~Beavis\bnl, 
  C.~Besliu\bucharest, 
  B.~Budick\newyork, 
}
\author{
  H.~B{\o}ggild\nbi, 
  C.~Chasman\bnl, 
  C.~H.~Christensen\nbi, 
}
\author{
  P.~Christiansen\nbi,  
  R.~Debbe\bnl, 
  E. Enger\oslo,  
}
\author{
  J.~J.~Gaardh{\o}je\nbi, 
  M.~Germinario\nbi, 
  K.~Hagel\texas, 
}
\author{
  A.~Holm\nbi, 
  H.~Ito\bnl$^,$\kansas, 
  A.~Jipa\bucharest, 
  F.~Jundt\ires,  
  J.~I.~J{\o}rdre\bergen, 
}
\author{
  C.~E.~J{\o}rgensen\nbi, 
  R.~Karabowicz\krakow, 
  E.~J.~Kim\bnl, 
  T.~Kozik\krakow, 
}
\author{
  T.~M.~Larsen\oslo, 
  J.~H.~Lee\bnl, 
  Y.~K.~Lee\baltimore, 
  S.~Lindal\oslo, 
  G.~Lystad\bergen,  
}
\author{
  G.~L{\o}vh{\o}iden\oslo, 
  Z.~Majka\krakow, 
  A.~Makeev\texas, 
}
\author{
  M.~Mikelsen\oslo, 
  M.~Murray\texas$^,$\kansas, 
  J.~Natowitz\texas, 
}
\author{
  B.~S.~Nielsen\nbi, 
  D.~Ouerdane\nbi, 
  R.~P\l aneta\krakow, 
}
\author{
  F.~Rami\ires, 
  C.~Ristea\nbi, 
  O.~Ristea\bucharest, 
  D.~R{\"o}hrich\bergen, 
}
\author{
  B.~H.~Samset\oslo, 
  D.~Sandberg\nbi, 
  S.~J.~Sanders\kansas, 
}
\author{
  P.~Staszel\nbi$^,$\krakow, 
  T.~S.~Tveter\oslo, 
  F.~Videb{\ae}k\bnl, 
  R.~Wada\texas, 
  H.~Yang\bergen, 
}
\author{
  Z.~Yin\bergenc ~and
  I.~S.~Zgura\bucharest 
}
\end{center}
\author{
  (The BRAHMS Collaboration) 
}

\address{
  \bnl~Brookhaven National Laboratory, Upton, New York 11973, USA\\
  \ires~Institut de Recherches Subatomiques and Universit{\'e} Louis
  Pasteur, Strasbourg, France\\
  \kraknuc~Institute of Nuclear Physics, Krakow, Poland\\
  \krakow~M. Smoluchkowski Institute of Physics, Jagiellonian University,
  Krakow, Poland\\
  \baltimore~Johns Hopkins University, Baltimore 21218, USA\\
  \newyork~New York University, New York 10003, USA\\
  \nbi~Niels Bohr Institute, Blegdamsvej 17, University of Copenhagen,
  Copenhagen 2100, Denmark\\
  \texas~Texas A$\&$M University, College Station, Texas, 17843, USA\\
  \bergen~University of Bergen, Department of Physics, Bergen,
  Norway\\
  \bucharest~University of Bucharest, Romania\\
  \kansas~University of Kansas, Lawrence, Kansas 66045, USA \\
  \oslo~University of Oslo, Department of Physics, Oslo, Norway\\
 }

\corauth[a]{Corresponding author.\\
{\it Email address:} Zhongbao.Yin@ift.uib.no}

\begin{abstract}
  We present spectra of charged pions and protons in 0-10\% 
  central Au+Au collisions
  at $\sqrt{s_{NN}}=200$ GeV at mid-rapidity ($y=0$) and forward pseudorapidity 
  ($\eta=2.2$) measured with the BRAHMS experiment at
  RHIC.  The spectra are compared
  to spectra from p+p collisions at the same
  energy scaled by the number of binary collisions.  The resulting
  nuclear modification factors for central Au+Au collisions
  at both $y=0$ and $\eta=2.2$ 
  exhibit suppression for charged pions but not for 
  (anti-)protons at intermediate $p_T$. 
  The $\bar{p}/\pi^-$ ratios have been measured up to $p_T\sim 3$ GeV/$c$ 
  at the two rapidities and the results indicate 
  that a significant fraction of the charged 
  hadrons produced at intermediate $p_T$ range are (anti-)protons at both 
  mid-rapidity and $\eta = 2.2$.

\end{abstract}

\begin{keyword}
Particle production \sep Nuclear modification factor

\PACS 25.75 Dw
\end{keyword}
\end{frontmatter}


\section{Introduction}

One of the reasons for studying heavy-ion collisions at high energies 
is to search 
for the predicted Quark-Gluon Plasma (QGP), a deconfined state of 
quarks and gluons, and to investigate the properties of this 
state of matter at extremely high energy densities. 
High $p_T$ hadrons, primarily produced from the fragmentation of the 
hard-scattered partons, are considered a good probe of 
the QGP~\cite{Bjorken83,Gyulassy90,Baier95}.
Due to induced gluon radiation, hard-scattered partons will suffer 
a larger energy loss in a hot dense medium of color
charges than in color neutral matter. This results in fewer 
charged hadrons produced at moderate to high $p_T$; the hadrons 
are said to be suppressed. Indeed, all four
experiments at RHIC have observed that high $p_T$ inclusive hadron yields
in central Au+Au collisions are suppressed as compared to p+p and d+Au 
interactions at mid-rapidity~\cite{BRAHMShpt,PHENIXhpt,PHOBOShpt,STARhpt}. 
However, it was also discovered that  
the yields of protons and anti-protons 
at intermediate $p_T$ (1.5-4.5 GeV/$c$) are 
comparable to those of pions and not suppressed 
at mid-rapidity as compared to elementary 
nucleon-nucleon collisions~\cite{PHENIXhpt,PHENIX_06}. 
These experimental results have motivated several suggestions on 
how hadrons are produced at intermediate 
$p_T$~\cite{Vitev03,HwaYang03,Fries03,Greco03}, 
such as the possibility that boosted quarks from a collectively 
expanding QGP recombine to form the final-state 
hadrons~\cite{HwaYang03,Fries03,Greco03}. 
Among the interesting results 
from the BRAHMS experiment is that
at forward pseudorapidity $\eta = 2.2$ inclusive negatively
charged hadrons are suppressed in both central Au+Au and minimum-bias
d+Au collisions~\cite{BRAHMShpt,BRAHMSdAu}. This raises the possibility
that initial-state effects such as gluon saturation
may also influence hadron production 
at intermediate $p_T$~\cite{Khar03,Baie03,Jali03}.

To explore the effect of the nuclear medium on intermediate 
$p_T$ particle production, we present in this paper the 
invariant $p_T$ spectra of charged pions and 
protons measured by the BRAHMS experiment at RHIC
up to 3 GeV/$c$ in central Au+Au collisions at 
$\sqrt{s_{NN}} = 200$ GeV at both mid-rapidity and forward pseudorapidity 
$\eta = 2.2$.  The spectra are then compared
to reference data from p+p collisions
at the same energy scaled by the number of binary collisions 
$\langle N_{bin}\rangle$ by using the nuclear modification factor:
\begin{equation}
R_{AA}=\frac{d^2N^{AA}/dp_Tdy}
{({\langle N_{bin}\rangle}/\sigma_{inel}^{pp})d^2\sigma^{pp}/dp_Tdy} ,
\end{equation}

\noindent
where $d^2N^{AA}/dp_Tdy$ 
is the differential yield per event in the nucleus-nucleus (A+A) collision, 
and $\sigma_{inel}^{pp}$ and $d^2\sigma_{inel}^{pp}/dp_Tdy$ are 
the total and differential cross section for inelastic p+p collisions, 
respectively.

\section{Experiment and data analysis}

The BRAHMS experiment~\cite{BRAHMSNIM} 
consists of event characterization detectors and 
two independent magnetic spectrometers, the 
mid-rapidity spectrometer (MRS) and the Forward Spectrometer (FS), 
both of which can be rotated in the horizontal plane around 
the beam direction. 
For the present 
studies the MRS was positioned at 90 degrees and the FS at 12 degrees with 
respect to the beam direction. Collision centrality is determined from 
the charged particle multiplicity measured by multiplicity detectors as 
described in~\cite{BRAHMSmult}. The trajectories of charged particles 
are reconstructed in the tracking devices (time projection chambers 
and drift chambers). The resulting straight-line track segments in two 
detectors located on either side of a magnet are then matched and 
the particle momentum is determined from the deflection of
the track in the magnetic field. The intrinsic momentum resolution of 
the spectrometers
at maximum magnetic field setting is $\delta p/p = 0.0077p$ for the MRS and
$\delta p/p = 0.0008p$ for the FS~\cite{BRAHMSdAu}, 
where $p$ is written in units of GeV/$c$.
In the MRS charged particles are identified
using a time-of-flight wall (TOFW), whereas in the FS a time-of-flight 
wall (H2) and a ring imaging Cherenkov (RICH) detector are used for particle 
identification (PID). To identify charged pions and protons using 
time-of-flight detectors, $2\sigma$ standard deviation PID cuts in the derived
$m^2$ and momentum space are imposed for each species. With a 
timing resolution of $\sigma_{\rm{TOFW}} \sim 80$ ps in the Au+Au runs, 
protons and pions can be well separated from
kaons up to momenta 3.2 GeV/$c$ and 2.0 GeV/$c$, respectively.
For pions above 2 GeV/$c$, an asymmetric PID cut is applied, {\it i.e.} the
region where the pion and kaon $2\sigma$ cuts overlap is excluded for PID, 
and the pion yield in the
region is obtained by assuming a symmetric PID distribution about the mean 
pion mass squared value. This allows the pion $p_T$ spectrum to be 
extended to 3 GeV/$c$, at which point the kaon contamination of 
pions is estimated to be less than 5\% and is accounted for 
in the systematic errors. For the FS PID in 
the present analysis, H2 is used only for the low momentum data. 
With a timing resolution of 
$\sigma_{\rm{H2}} \sim 90$ ps,
protons and pions can be identified up to 7.1 GeV/$c$ and 4.2 GeV/$c$,
respectively with a $2\sigma$ separation. 
Above 7.1 GeV/$c$, an asymmetric PID cut is applied and 
the proton yields in the overlap region
are estimated by assuming a symmetric PID distribution about the mean proton mass
squared value. Between 7.9 GeV/$c$ and 9 GeV/$c$, 
the Cherenkov threshold for protons, the RICH detector is used to
determine the kaon contamination of the proton spectrum. At 9 GeV/$c$ the 
contamination of protons by kaons is estimated to be less than 6\%. 
Above 9 GeV/$c$, protons are identified by using the RICH to veto pions and
kaons. To identify pions, the RICH is directly used to separated pions from 
kaons well from momentum of 2.5 GeV/$c$ up to 20 GeV/$c$. 

The invariant differential yields 
$\frac{1}{2\pi}\frac{d^2N}{p_Tdydp_T}$ 
(respectively $\frac{1}{2\pi}\frac{d^2N}{p_Td{\eta}dp_T}$ 
at forward rapidity)   
were constructed for each spectrometer 
setting. As discussed in~\cite{BRAHMSstopping}
the differential yields were corrected for geometrical acceptance, 
tracking and PID inefficiencies, in-flight decay of pions, 
the effect of absorption and multiple scattering. 
The pion contamination by hyperon ($\Lambda$) and
neutral $K^0_{S}$ decays were investigated in~\cite{BRAHMSmeson} and found
to be less than 5\% in the MRS and 7\% in the FS, respectively. The 
contribution to proton spectra by the $\Lambda$ decays was estimated with 
a GEANT~\cite{GEANT} simulation where an exponential distribution in $p_T$ 
with inverse slope taken from the PHENIX and STAR 
measurements~\cite{PHENIXlambda,STARlambda} for both (anti-)protons and
(anti-)lambdas was generated for several spectrometer settings. By taking
the ratio of $\Lambda(\bar{\Lambda})$ to $p(\bar{p})$ yields of 
0.89 (0.95)~\cite{PHENIXlambda} 
in 0-10\% central Au+Au and 0.45 (0.55)~\cite{STARlambda} 
in p+p collisions at $\sqrt{s_{NN}} = 200$ GeV and 
assuming a constant behavior in the rapidity
interval $|y| \leq 2.2$ as indicated by HIJING model~\cite{HIJING}, it is 
found that the fraction of protons originating from $\Lambda(\bar{\Lambda})$ 
decays is at a maximum value around 35-40\% in central Au+Au 
and 27-30\% in p+p collisions and decreases with $p_T$. In the 
following correction for feed-down from the
(anti-)lambda decays has been applied, whereas the contamination of pions due 
to weak decays has not been corrected but is accounted for in the systematic 
errors.

\section{Particle spectra}

\begin{figure}[htp]
\vspace{2mm}
  \epsfig{file=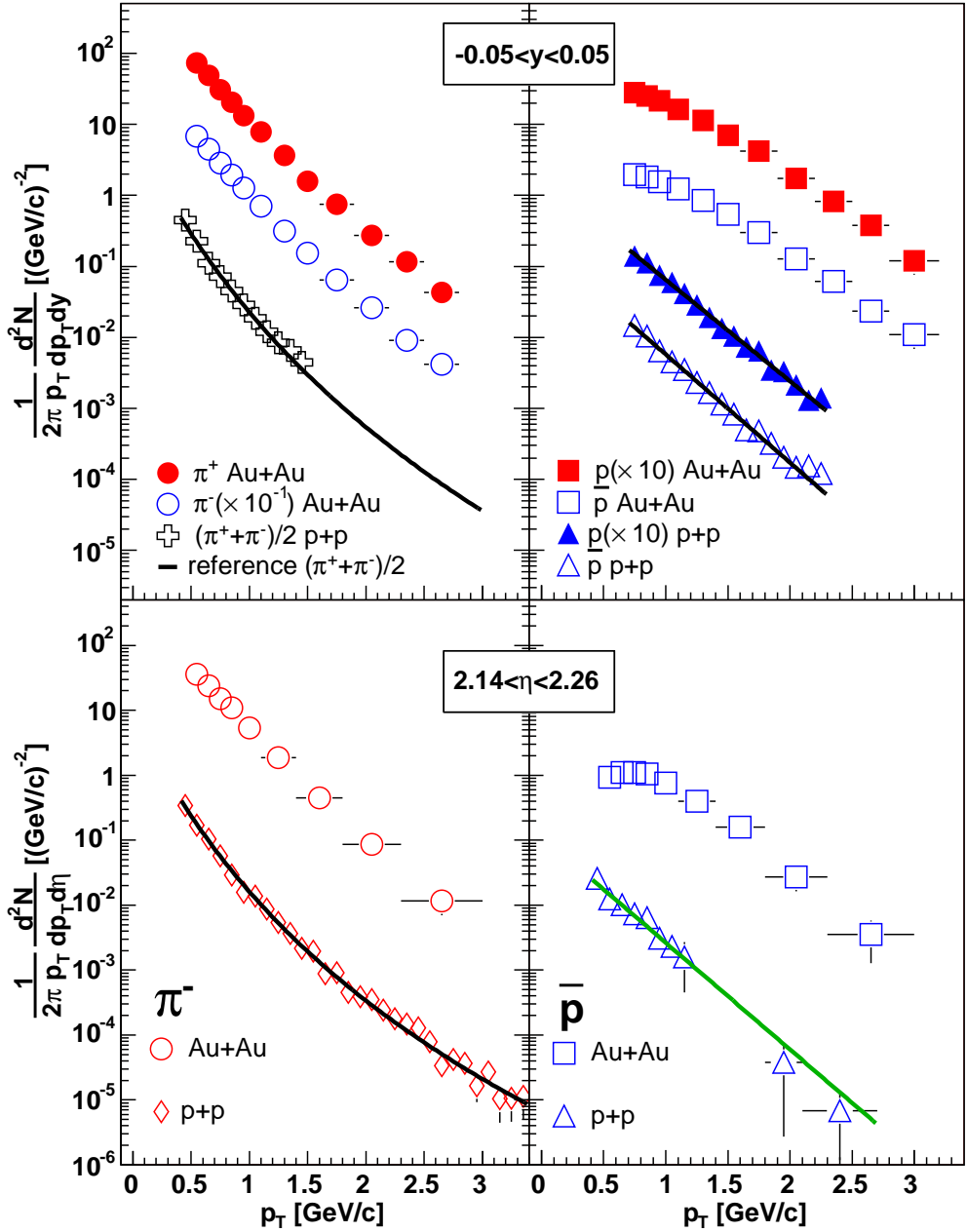,width=\linewidth}
  \caption{ Top row: $p_T$ spectra of charged pions (left panel) and 
protons (right panel) at mid-rapidity in 0-10\% central Au+Au and p+p 
collisions at $\sqrt{s_{NN}}$ = 200 GeV. 
The error bars are statistical only. The systematic errors
are estimated to be less than 15\% for pions and 18\% for (anti-)protons.
For the reference spectrum the systematic error is estimated to be less 
than 19\%. For clarity, some spectra are scaled vertically as quoted.
Bottom row: $p_T$ spectra for $\pi^-$ and $\bar{p}$ at forward rapidity
$\eta = 2.2$ in 0-10\% central Au+Au 
and p+p collisions at $\sqrt{s_{NN}} = 200$ GeV.
 The systematic errors are estimated to be
14\% for pions and 17\% for anti-protons.}
  \label{spectra}
\end{figure}

The top row of Figure~\ref{spectra} shows the $p_T$ spectra of 
charged pions (left panel)
and protons (right panel) at mid-rapidity in 0-10\% central Au+Au and p+p 
collisions at $\sqrt{s_{NN}} = 200$ GeV. Also shown in the left panel of 
the figure is the measured spectrum of $(\pi^++\pi^-)/2$ in p+p collisions,
where pions can only be identified up to 1.5 GeV/$c$ with 
TOFW for 2003 p+p runs. We thus constructed a reference spectrum shown as a solid line 
by dividing the neutral pion spectrum in p+p collisions measured by 
PHENIX~\cite{PHENIXpi0} by the spectrum from 
PYTHIA simulation~\cite{PYTHIA} at the same rapidity range 
and then multiplying the results by the $(\pi^++\pi^-)/2$ spectrum
from PYTHIA. 
The spectra of (anti-)protons in 
p+p collisions are measured by the BRAHMS spectrometer but 
to a smaller $p_T$ coverage compared to those in Au+Au collisions due to 
a worse TOF resolution in the p+p runs. The spectra have 
been corrected for the trigger inefficiency~\cite{BRAHMSdAu} and fitted with 
an exponential function as shown in solid lines with the rapidity density and 
the inverse slope 
parameter of $0.101\pm 0.004$ ($0.098\pm 0.004$) and 
$0.304\pm 0.005$ ($0.285\pm 0.005$ GeV) for proton (anti-proton), 
respectively. The error bars are statistical only. The systematic errors in 
the measured spectra, which come from the uncertainties in the momentum 
determination, the time-of-flight measurements and ring radius reconstruction 
procedures, and the uncertainties in the corrections estimations, are estimated
to be less than 15\% for pions and 18\% for (anti-)protons. 
The systematic error in the reconstructed reference 
spectrum for charged pions is estimated to be less than 19\%.

The bottom row of Figure~\ref{spectra} shows the $p_T$ spectra 
for $\pi^-$ (left panel) and $\bar{p}$ (right panel) 
at forward rapidity $\eta = 2.2$ in 0-10\% central Au+Au and
p+p collisions at $\sqrt{s_{NN}} = 200$ GeV. Solid lines
are curves fit to the $\pi^-$ and $\bar{p}$ spectra
in p+p collisions. The spectra
are constructed in terms of  $d^2N/dp_Td\eta$ because the rapidity coverages
of the FS at 12 degrees for pions and protons are different, making a 
comparison of anti-proton to pion yields difficult. In addition, since
the Jacobian effect is largest at mid-rapidity and gets rather small at
larger rapidities at intermediate $p_T$ range as we focused on in this paper, 
we expect the conclusions drawn from the spectra expressed in terms
of $d^2N/dp_Td\eta$ should be the same as from those of $d^2N/dp_Tdy$.

\begin{figure}[htp]
  \epsfig{file=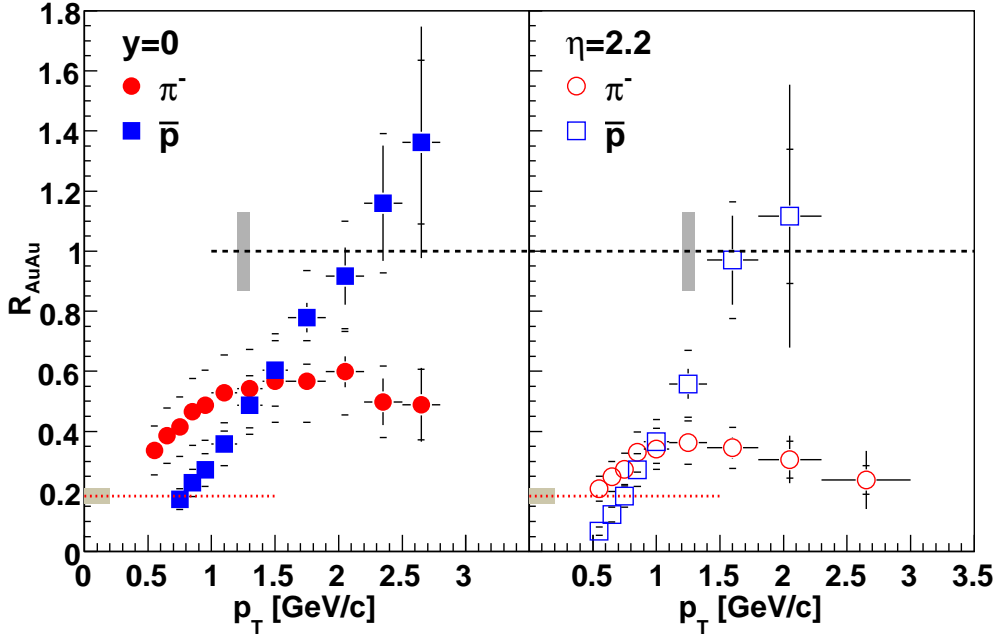,width=\linewidth}
 \vspace{-5mm}
  \caption{Nuclear modification factors for $\pi^-$ and $\bar{p}$
    measured for 0-10\% central Au+Au collisions
    at $\sqrt {s_{NN}} = 200$ GeV at mid-rapidity (left panel) and
    $\eta = 2.2$ (right panel).  Error bars represent statistical
    errors; the systematic errors are indicated by horizontal lines.  
    The dotted and dashed lines indicate the expectations of
    participant scaling and binary scaling, respectively.
    The shaded bars represent the systematic errors associated with
    the determination of these quantities.
    Systematic errors other than the uncertainties in
    $\langle N_{bin}\rangle$ determinations are estimated to be 20\%
    except for $\pi^-$ at mid-rapidity, where they are around 24\%.}
  \label{Raa}
\end{figure}

\section{Nuclear modification factor}

In Figure~\ref{Raa} the nuclear modification factors 
for $\pi^-$ and $\bar{p}$ are deduced for 0-10\% central Au+Au collisions
at $\sqrt {s_{NN}} = 200$ GeV at mid-rapidity (left panel) and
$\eta = 2.2$ (right panel). Error bars represent statistical
errors; the systematic errors are indicated by horizontal lines.
The dotted and dashed lines indicate the expectations of
participant scaling and binary scaling, respectively.
The shaded bars represent the systematic errors associated with
the determination of these quantities. Systematic errors 
other than the uncertainties in
$\langle N_{bin}\rangle$ determinations are estimated to be 20\% except
for $\pi^-$ at mid-rapidity, where they are around 24\%. 

\begin{figure}[htp]
  \epsfig{file=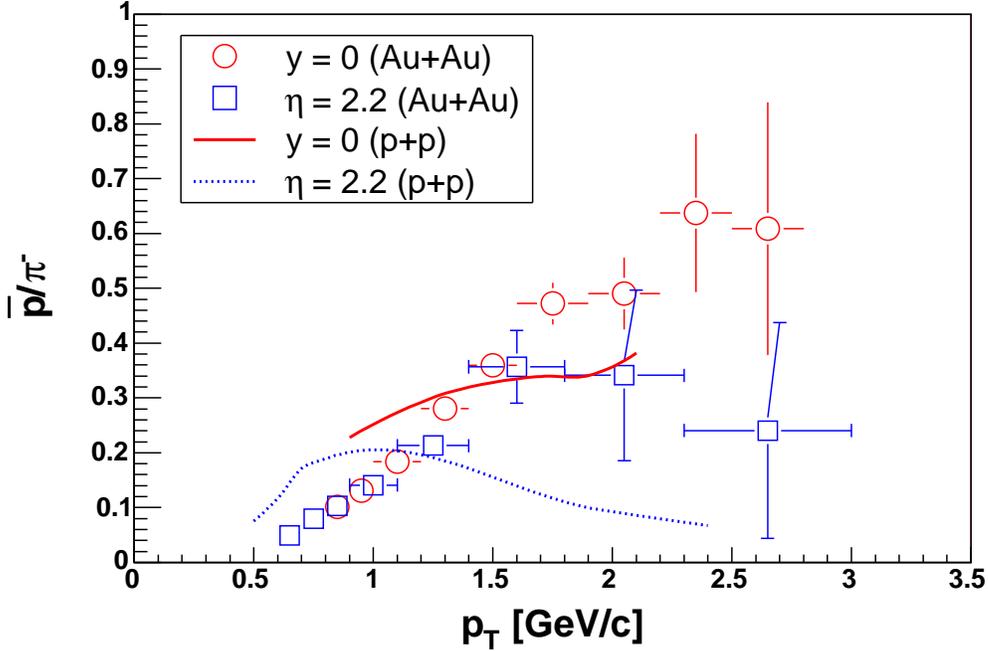,width=\linewidth}
 \vspace{-5mm}
  \caption{$\bar{p}/\pi^-$ ratios at both mid-rapidity and $\eta =2.2$ for
    0-10\% central Au+Au collisions at $\sqrt{s_{NN}} = 200$ GeV. The error
    bars show the statistical errors only. The systematic errors are estimated
    to be less than 12\% at both $y=0$ and $\eta = 2.2$. The corresponding ratios 
    in $p+p$ collisions at $\sqrt{s_{NN}} = 200$ GeV are sketched as solid line
    and dotted line, respectively.
  }
  \label{pPiRatio}
\end{figure}

Similar to the unidentified charged hadrons~\cite{BRAHMShpt} at both
mid-rapidity and forward rapidity, $R_{AuAu}$
for charged pions increases monotonically up to $\sim 1.5$ GeV/$c$ 
and levels off at a value below unity above 1.5 GeV/$c$ 
indicating that charged pions yields
are suppressed with respect to p+p collisions at intermediate $p_T$. 
Furthermore, the $\pi^-$ yields at forward rapidity 
show a similar or even stronger suppression, indicating that nuclear effects 
other than parton energy 
loss (jet quenching) might be contributing 
to the strong suppression. 
The suppression at midrapidity around $p_T$ $\sim 2$ GeV/$c$ is smaller 
(about 30\%)
than the suppression that has been reported for neutral pions 
\cite{PHENIXpi0_AuAu} and which is seen at forward rapidity. This
difference can - to a large extent - be attributed to the construction 
of the reference spectrum
and has been accounted for by 
the systematical error for $\pi^-$ at mid-rapidity. 
Another interesting feature 
shown in the figure is that the anti-proton
yields at both mid-rapidity and $\eta = 2.2$ are not suppressed at 
$p_T > 1.5$ GeV/$c$. 

\section{Particle ratios}

Figure~\ref{pPiRatio} shows $\bar{p}/\pi^-$ ratios at both mid-rapidity 
and $\eta =2.2$ for 0-10\% central Au+Au collisions 
at $\sqrt{s_{NN}} = 200$ GeV. The error
bars show the statistical errors only. 
In the present ratios, most systematic errors cancel out. Remaining systematic
errors arising from PID efficiencies, acceptance corrections,
corrections for nuclear interactions with detector etc. are estimated
to be less than 12\% at both $y=0$ and $\eta = 2.2$. 
Also shown in the figure are the corresponding ratios for p+p collisions 
at $\sqrt{s_{NN}} = 200$ GeV. There is a clear
increase of the $\bar{p}/\pi^-$ ratios at intermediate $p_T$
in central Au+Au collisions relative
to the level seen in p+p collisions (see also ~\cite{PHENIX_06,STAR_06}). 
This enhancement is most 
likely due to the interplay of several 
final-state effects and possibly a new hadronization mechanism other than 
parton fragmentation. Calculations based on a parton
recombination scenario~\cite{Greco03} with a collective flow at the
partonic level appear to be able to qualitatively describe the
data at mid-rapidity.    

\section{Summary}

In summary, the BRAHMS measurements demonstrate a significant
suppression of charged pions at intermediate $p_T$ 
at both mid-rapidity and forward rapidity for 0-10\% central 
Au+Au collisions at $\sqrt{s_{NN}}=200$~GeV. 
Such a strong suppression is believed to be 
caused primarily by the parton losing energy when traversing the
partonic (i.e. characterized by color degrees of freedom) 
medium created in central Au+Au collisions. 
The strong $\pi^-$ suppression at forward rapidity suggests 
that the hot dense partonic medium may also exist in the forward
rapidity region and that there might be other nuclear effects such as 
gluon saturation contributing to the suppression. 
However, the suppression is not observed for 
(anti-)protons at intermediate $p_T$ at either mid-rapidity or forward
pseudorapidity $\eta = 2.2$. $\bar{p}/\pi^-$ ratios in central Au+Au show an
enhancement of (anti-)proton production relative to the p+p collisions at intermediate
$p_T$. All these observations
are consistent with a picture where 
a dense strongly interacting partonic 
matter with a strong collective flow is most likely formed in central Au+Au collisions 
over a large rapidity range which results in the strong suppression of 
charged pion yields and boosts the protons to higher transverse momentum. 

This work was supported by the Division of Nuclear Physics of the
Office of Science of the U.S. DOE under contract DE-AC02-98-CG10886, 
the Danish Natural Science
Research Council, the Research Council of Norway, the Polish State
Commission for Scientific Research and the Romanian Ministry of
Research.

\end{document}